\documentclass[a4paper]{article}

\usepackage[english]{babel}
\usepackage[utf8x]{inputenc}
\usepackage[T1]{fontenc}
\usepackage{cite}
\usepackage{multirow}

\usepackage[a4paper,top=3cm,bottom=2cm,left=3cm,right=3cm,marginparwidth=1.75cm]{geometry}

\usepackage[
    type={CC},
    modifier={by},
    version={4.0},
]{doclicense}

\usepackage{amsmath}
\usepackage{amssymb}
\usepackage{graphicx}
\usepackage[colorinlistoftodos]{todonotes}
\usepackage[colorlinks=true, allcolors=blue]{hyperref}
\usepackage{hyperref}
\newcommand{\etal}{\textit{et al}.}
\newcommand{\etalspace}{\textit{et al}. }
\newcommand{\ie}{\textit{i}.\textit{e}. }
\newcommand{\eg}{\textit{e}.\textit{g}. }

\usepackage{parskip}

\usepackage{dsfont} 

\usepackage{setspace}
\title{Infectious Disease Forecasting for Public Health
\footnote{This chapter will appear as part of the book {\em Population Biology of Vector-Borne Diseases}, Eds. John Drake, Mike Strand, Mike Bonsall.}}
\author{Stephen A Lauer, Alexandria C Brown, Nicholas G Reich}

\begin{document}
\maketitle

\begin{abstract}
Forecasting transmission of infectious diseases, especially for vector-borne diseases, poses unique challenges for researchers.
Behaviors of and interactions between viruses, vectors, hosts, and the environment each play a part in determining the transmission of a disease.
Public health surveillance systems and other sources provide valuable data that can be used to accurately forecast disease incidence. 
However, many aspects of common infectious disease surveillance data are imperfect: cases may be reported with a delay or in some cases not at all, data on vectors may not be available, and case data may not be available at high geographical or temporal resolution.
In the face of these challenges, researchers must make assumptions to either account for these underlying processes in a mechanistic model or to justify their exclusion altogether in a statistical model.
Whether a model is mechanistic or statistical, researchers should evaluate their model using accepted best practices from the emerging field of infectious disease forecasting while adopting conventions from other fields that have been developing forecasting methods for decades.
Accounting for assumptions and properly evaluating models will allow researchers to generate forecasts that have the potential to provide  valuable insights for public health officials.
This chapter provides a background to the practice of forecasting in general, discusses the biological and statistical models used for infectious disease forecasting, presents technical details about making and evaluating forecasting models, and explores the issues in communicating forecasting results in a public health context.
\end{abstract}

\vfill

\doclicenseThis

\clearpage 

\tableofcontents

\clearpage

\begin{center}
``... diviners employ art, who, having learned the known by observation, \\ seek the unknown by deduction.''  \\ 
-- Cicero (44 BCE, as quoted in \cite{mccloskey1992art})

\ \\

``We may regard the present state of the universe as the effect of its past and the cause of its future. An intellect which at a certain moment would know all forces that set nature in motion, and all positions of all items of which nature is composed, if this intellect were also vast enough to submit these data to analysis, it would embrace in a single formula the movements of the greatest bodies of the universe and those of the tiniest atom; for such an intellect nothing would be uncertain and the future just like the past would be present before its eyes.''\\ -- LaPlace (1825, as quoted in \cite{Silver2012})

\end{center}

\section{Background}
\subsection{A brief history of forecasting}
The ability to foretell, or divine, future events for millennia has been seen as a valued skill.
While there are records of Babylonians attempting to predict weather patterns as early as 4000 BCE based on climatological observations \cite{Milham1918}, early attempts at divination were just as likely to be driven by unscientific observation. 
However, in the last 150 years, rapid technological advancements have made data-driven forecasting a reality across a number of scientific and mathematical fields. .

The science of forecasting was pushed forward especially in the second half of the 20th century by the fields of meteorology and economics, but more recently other fields have started to build on this research.
Examples include world population projections \cite{gerland2014world, raftery2012bayesian}, political elections,\cite{campbell1996polls, graefe2015german, lewis2014us}, seismology\cite{field2009uniform,bray2013assessment,Chambers2012}, as well as infectious disease epidemiology \cite{biggerstaff2016results, lowe2014dengue, viboud2017rapidd, held2017probabilistic, Reich2016a}.

\begin{figure}[htbp]
\begin{center}
\includegraphics[width=\textwidth]{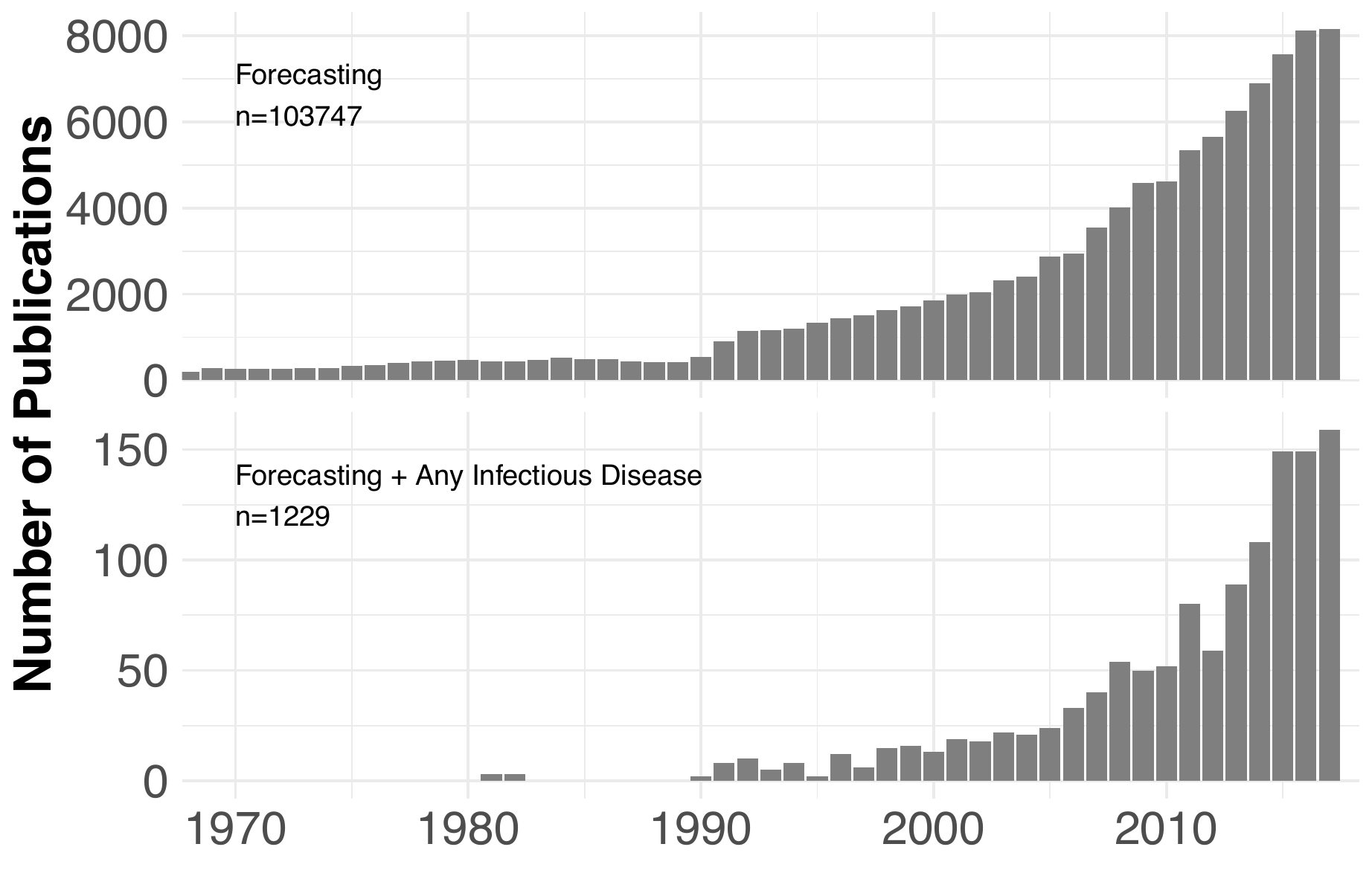}
\caption{Publication trends from 1970 through 2016. The y-axis in each panel shows the number of publications, according to the Web of Science, for papers with the topic of (A) `forecast*', and (B) `forecast*' + any of a list of infectious diseases taken from WHO \cite{WHO-all-id}. There are 1,989 and 0 publications, respectively, that were published prior to 1970. All counts are taken from the Science Citation Index and the Social Science Citation Index, obtained via the Web of Science database.}
\label{fig:publications}
\end{center}
\end{figure}

Forecasting has been an active and growing area of research for over a century (\autoref{fig:publications}), with particular acceleration observed since 1980.
While research focused on forecasting infectious diseases started in earnest in the 1990s, since 2005 the number of articles on infectious disease forecasting has increased seven-fold, at a faster pace than research on general forecasting during that time, which increased by a factor of 3.
In 1991, forecasting was the topic of one of every thousand published academic papers, based on counts from the Science Citation Index and the Social Science Citation Index, obtained from the Web of Science.
In 2017, over four of every thousand indexed publications were about forecasting.

\subsection{What is a forecast?}\label{sec:what}

In common parlance, there is not a strong distinction between the terms `prediction' and `forecast.'
Nor does there exist a strong consensus in the biomedical, ecological, or public health literature on the distinction.
Nate Silver has suggested that etymologically, the term forecast ``implied planning under conditions of uncertainty'' in contrast to prediction, which was a more ancient idea associated with superstition and divination \cite{Silver2012}.
In the modern scientific world, some fields, such as seismology, use the term forecast to refer to a probabilistic statement in contrast to a prediction which is a ``definitive and specific'' statement about a future event.
In other fields, the difference in meaning is even less clearly defined, with forecasting often connoting the prediction of a future value or feature of a time-series \cite{Diebold2001}.
We note that conventionally in biomedical research the term `prediction' refers to an individual-level clinical outcome (\eg risk-scores give individualized predictions of heart attack or stroke risk), but both `prediction' and `forecast' are used interchangeably to refer to events or outcomes that may impact more than one person.
The term `forecast' often refers broadly to a quantitative estimate about a future trend or event observable or experienced by many individuals as opposed to than a local, individualized outcome. 
Our use of the term forecasting takes aspects from several of these definitions.
Specifically, we define a forecast as
{\em a quantitative statement about an event, outcome, or trend that has not yet been observed, conditional on data that has been observed}.

Probabilistic forecasts represent an important class of forecasts. They are forecasts that have a statement about the uncertainty surrounding an event, outcome, or trend, and are not just providing a single ``best guess'' estimate about the event. For example, a ``point'' forecast for the number of dengue hemorrhagic fever cases in Singpore in the 14th calendar week of 2024 could be simply the number 25. However, a probabilistic forecast for the same target might say that while 25 cases is the most likely specific outcome, the probability of seeing exactly 25 cases is only 0.10 or 10\%.

Note that a forecasted event or outcome need not necessarily be in the future, as events in the past that have not yet been quantitatively measured may also be forecasted.
For example, on May 1 we may have data for a particular time-series available through April 1.
We could make a ``forecast'' of the time-series for April 15 even though this event is in the past.
This type of forecast has been referred to as a ``nowcast'',\cite{brooks2018nonmechanistic,reich2019collaborative} although more generally, this is a special case of a forecast.

\subsection{Forecasting challenges that are specific to infectious disease} \label{sec:challenges}

There are operational and statistical challenges in forecasting that are specific to the setting of infectious disease.
These challenges in and of themselves may not be unique to the field, but taken together, they describe obstacles that forecasters face when taking on a problem in infectious disease.

\subsubsection*{Challenge 1: System Complexity}

When attempting to forecast the transmission of an infectious disease, and in particular a vector-borne disease, researchers need to account for processes on micro and macro scales.
Behaviors of and interactions between viruses, vectors, hosts, and the environment each play a part in determining the transmission of a disease.
For example, it has been hypothesized that rapid changes in climate may lead to unforeseen shifts in vector population dynamics, highlighting the fragility of models that attempt to generate forecasts based on existing knowledge about these complex systems.\cite{mandal2011mathematical,jewell2015bayesian,ostfeld2015climate} 

Researchers have developed mechanistic models based on biological and behavioral principles that encode the processes by which diseases spread (Section \ref{sec:mech}).
For vector-borne diseases, these models often include complex dynamics (see, e.g. \cite{mandal2011mathematical} and \cite{reiner2013systematic}).
That said, vector dynamics are often omitted from models of vector-borne disease for simplicity and tractability.\cite{reiner2013systematic}
While numerous mechanistic models have been fit to data to provide inference about disease transmission parameters for vector-borne diseases \cite{lourencco20142012,reiner2013systematic}, far fewer studies have examined prospective forecasting performance of such models \cite{Defelice2017,jewell2015bayesian}.

The need for models to mirror the ecological complexity of the system stands in conflict with a fundamental principle of forecasting which is to use as simple a model as necessary.
Providing a good `fit' to in-sample data will not guarantee that such a model will generate accurate or even reasonable forecasts.
Due to the dearth of high-resolution data on vector populations and host infections, understanding whether and how such detailed biological data can improve the accuracy of forecasts of population-level transmission largely remains to be seen (Section \ref{sub:lab}).

\subsubsection*{Challenge 2: Data Sparsity}

A central challenge in forecasting vector-borne disease, and infectious diseases more generally, is how to balance the complexity of the biological and social models used with the coarseness of available data \cite{moran2016epidemic}.
For forecasting weather, scientists rely on tens of thousands of sensors across the world, collecting continuous real-time data. 
There are no analogues to these rich and highly accurate data streams for infectious disease researchers.
The gold-standard data in epidemiological surveillance arise from systems that typically capture only a fraction of all cases and often are reported with substantial delays and/or revisions to existing data.
Technology shows some promise to provide more richly detailed data in a timely fashion about humans, climate, and vectors alike.\cite{george2019}
However, new methods and data streams will be required to develop and implement increasingly refined forecasting models for vector-borne diseases.

\subsubsection*{Challenge 3: The Forecasting Feedback Loop}

Forecasts of disease incidence can encourage governments and public health organizations to intervene to slow transmission.
If forecasts of infectious disease are used to inform targeted interventions or risk communication strategies and the interventions change the course of the epidemic, then the forecast itself becomes enmeshed in the causal pathway of an outbreak. 
This feedback loop has been identified as the single most important challenge separating infectious disease forecasting from forecasting natural phenomena such as weather \cite{moran2016epidemic}.

In settings where forecasts will be used to inform interventions, this feedback loop of infectious disease forecasting should be taken into account in the forecasts. 
Without such accounting, if a forecast predicts an outbreak and triggers an intervention that prevents the epidemic from occurring, then the forecast itself would be seen as wrong, despite this being a public health victory. 
This implies that forecasting models should, when in these settings, create multiple forecasts under different intervention scenarios. 
Mechanistic forecasting models, that use explicit disease transmission parameters, may be best suited for these types of forecasts, since intervention effects could be incorporated directly as impacting these parameters.
However, any forecasts from such a model should be subjected to intense scrutiny, since it will necessarily be based on very strong assumptions about the intervention and transmission patterns. 

Methodological development is needed in this area to address open scientific questions.
What model frameworks can best balance forecast accuracy with the ability to incorporate multiple potential future scenarios?
Can forecast models be used to assess intervention effectiveness?

\subsection{Definitions and basic notation}
\label{sec:notation}

Here, we introduce some basic mathematical notation for time-series forecasting that we use throughout this chapter.
In many forecasting applications, the available data are often a time series of observed values for a particular location or setting.
For infectious disease applications, these data are often a measure of incidence, such as case counts or the percentage of all doctor visits with primary complaint about a particular disease.
In the text that follows, we use language specific to that of spatio-temporal disease incidence data, although much of what we describe can be applied more generally as well.

\subsubsection*{Data}

We start with a simple example and later extend the notation to more realistic scenarios.
In our example, we have a complete (\ie no missing data) time series of infectious disease case counts from a single location, such as a school or hospital. 
We define $y_t$ as an observed value of this incidence in time interval $t$ from our time series $\{y_1, y_2, y_3, \dots\, y_t, \dots, y_T\}$.
We assume that these observations are draws from random variables $Y_1, Y_2, Y_3, \dots, Y_t, \dots, Y_T$, where the probability distribution of $Y_{t+1}$ may be dependent on $t$, prior values of $y$ represented as $y_{1:t}$, and a matrix of other covariates $\mathbf{x}_t$.
(Often, the analyst may wish to include multiple different lagged values of a single covariate. In this notation, for simplicity, these would all be considered to be part of $\mathbf{x}$.)
We use $T$ throughout to refer to the total number of time points in the time-series and $t$ to refer to a specific time point relative to which a forecast is generated.

Two important features of our observed data are frequency and scale.
In our example, incidence is recorded at regular time intervals.
Furthermore, many infectious disease time series have a cyclical element.
We define the frequency of a time series as the number of observations within a single cycle.
For example, if we have monthly incidence data and know that there are annual weather patterns that influence incidence in our observed data, the frequency of our time series would be 12 months or observations. 

\subsubsection*{Targets}

Targets are the as-yet-unknown features of the data that are the subject of forecasting.
In our example, we may want to forecast incidence at a certain future time---but targets can be a variety of endpoints extrapolated from the observed data.
For forecasts of the time-series values itself, \ie when a target is defined to be a past or future value of the time-series $Y_{t+k}$, we use a special nomenclature, referring to them as `k-step-ahead` forecasts.  
In general, we define $Z_{i|t}$ as a random variable for target $i$ positioned relative to time $t$. For example, in the infectious disease context, $Z_{i|t}$ could be:
\begin{itemize}
 	\item incidence at time $t$, or $Y_t$,
    \item incidence at time $t+k$ either in the future or past relative to time $t$, or $Y_{t+k}$, where $k$ is a positive or negative integer,
    \item peak incidence within some period of time or season, or $\max_{t\in \mathcal{S}} (Y_t)$ where $t$ are defined to be within season $\mathcal{S}$,
    \item the time(s) at which a peak occurs within some season, or $\{ t' \in \mathcal{S} : Y_{t'} = \max_{t\in \mathcal{S}} (Y_t) \} $
    \item a binary indicator of whether incidence at time $t+k$ is above a specified threshold $C$, or $\mathds{1} \{Y_{t+k}>C\}$.
\end{itemize}

\subsubsection*{Forecasts}

A forecast, as defined in Section \ref{sec:what}, must provide {\em quantitative} information about an outcome.
In the context of this notation, a probabilistic forecast can be represented as a predictive density function for a target, or $f_{z_{i|t}}(z|y_{1:t}, t, \mathbf{x}_{t})$.
The form of this density function depends on the type of variable that $Z$ is, and it could be derived from a known parametric distribution or specified directly.
For example, if the target is a binary outcome (\eg whether in week 4, the observed incidence will be above 10 cases) the density could be specified as a Bernoulli distribution with a parameter associated with the probability of the outcome occurring.
It could also be specified directly as a probability that the incidence is >10 and the probability that the incidence is $\leq 10$.
For an integer-valued target (\eg the number of new cases occurring in February), the predictive density could be represented by a Poisson distribution with a given mean or as a vector of probabilities associated with all possible integer values of cases.

To enable clear definitions for forecasting in real-time, forecasts must be associated with a specific time $t$.
This time $t$ represents the point relative to which targets are defined.
For example, if a forecast is associated with week 45 in 2013, then a `-1-step-ahead' (read `minus-one-step-ahead') forecast would be associated with incidence in week 44 of 2013 and a `3-step-ahead' forecast would be associated with week 48.

\subsubsection*{Forecast time-scale}

Another consideration for infectious disease forecasting is the forecast horizon, the temporal range that the forecast predicts \cite{Myers2000, Soyiri2013}.
Regardless of the model type, many recent infectious disease forecasting efforts have focused on short time scales (weeks or months) \cite{Birrell2011, Buczak2012, Gerardi2011, Hii2012, Lowe2011, Lu2010, Nishiura2011, Reich2016a, Shaman2013, Shaman2014, Sumi2012, Yan2010}.
These studies demonstrated the importance of recent case counts and seasonality on the immediate trajectory of infectious disease incidence.
In 2015, the National Oceanic and Atmospheric Administration (NOAA) and the Centers for Disease Control and Prevention (CDC) hosted a competition to make within-season forecasts for longer forecast horizons, such as annual dengue incidence, epidemic peak, and peak height, for San Juan, Puerto Rico and Iquitos, Peru \cite{NOAA2015}.
Prior to these competitions, long-term forecasts were more commonly used for chronic disease prevalence than for non-chronic infectious disease incidence \cite{Soyiri2013}.

\section{Models used for forecasting infectious diseases}

\subsection{Mechanistic vs. Statistical: a taxonomy of forecasting models}

According to Myers \etal, forecasting models for infectious diseases take either a `biological approach' or a `statistical approach' \cite{Myers2000}.
Others have phrased this distinction as one of mechanistic (\ie biological) and phenomenological (\ie statistical) models.
A model based on disease biology can account for previously unforeseen scenarios that are possible due to transmission dynamics, however these models often require specification of a large number of parameters and covariates in order to make forecasts.
On the other hand, statistical forecasting models are restricted by the assumption that future incidence will follow the patterns of incidence observed in the past, but can be specified without full knowledge of the disease process or interactions between members of the population.
In this section, we discuss the major modeling methods across the biological-statistical spectrum (\autoref{tab:methods}).

\begin{table}[htp]
\caption{
A summary of selected forecasting papers organized by vector-borne diseases (VBD) and other. Additionally, for each paper, we indicate which method is used, whether the method was mechanistic or statistical, whether the vector was specifically modelled, and the disease of interest.
}
\begin{center}
\begin{tabular}{l|l|p{1cm}|p{1.8cm}|l}
  & \textbf{Method} & \textbf{SIR model} & \textbf{Vector modelled?} & \textbf{Disease} \\
\hline

\multirow{12}{*}{VBD} 
 & EAKF & SI & Yes & West Nile \cite{Defelice2017} \\
 & Ordinary diff. eqs. & SEIR & Yes & dengue \cite{lourencco20142012}, malaria \cite{tompkins2019dynamical} \\
 & Ross-MacDonald NN & SIR & Yes & dengue \cite{Dinh2016} \\
 & Ross-MacDonald ODE & SIR & Yes & dengue \cite{Amaku2016,Zhu2018} \\
  & Stochastic diff. eqs. & SI & Yes & {\em Theileria orientalis} \cite{jewell2015bayesian} \\
& ARIMA & - & No & dengue \cite{Johansson2016}, malaria \cite{Anwar2016,Zinszer2015} \\

 & Cellular automaton & SIR & No & dengue \cite{Gerardi2011} \\
 & EAKF & SIR & No & dengue \cite{yamana2016superensemble} \\
 & GLM/Regression & - & No & dengue \cite{lauer2018prospective,Hii2012,lowe2014dengue,Reich2016a}, malaria \cite{Sewe2017}, other \cite{Haemig2011,Liu2014,Moore2012}  \\
 & Holt-Winters & - & No & dengue \cite{buczak2018ensemble} \\
 & KCDE & - & No & dengue \cite{Ray2017} \\
 & Neural networks & - & No & malaria \cite{Thakur2019}, Zika \cite{akhtar2018dynamic} \\
 \hline
\multirow{14}{*}{Other} 
 & EAKF & SEIRX & - & Ebola \cite{Shaman2014} \\
 & Markov chain & SEIR & - & Ebola \cite{gaffey2017application} \\
 & Stochastic diff. eqs. & SEIR & - & Ebola \cite{funk2016real, asher2018forecasting} \\
 & Agent-based & SEIR & - & influenza \cite{Hyder2013,Nsoesie2013} \\
 & Chain binomial & SEIR & - & influenza \cite{Nishiura2011} \\
 & Compartmental & SEIR & - & influenza \cite{Birrell2011} \\
 & Dynamic Bayesian & SIR & - & influenza \cite{osthus2017forecasting,Osthus2019} \\
 & EAKF & SIRS & - &  influenza \cite{Shaman2012,pei2017counteracting} \\
 & Expert opinion & - & - & influenza \cite{farrow2017human} \\
 & GLM/Regression & - & - & influenza \cite{Held2012,Goldstein2011} \\
 & Random forest & - & - & influenza \cite{Kane2014} \\
 & ARIMA & - & - & influenza \cite{Dugas2013}, other \cite{Yan2010,Lu2010} \\
 & Neural network & - & - & influenza \cite{xu2017forecasting}, hepatitis A  \cite{guan2004forecasting} \\
 & Holt-Winters & - & - & leprosy \cite{deiner2017short} \\
 \hline
\end{tabular}
\end{center}
\label{tab:methods}
\end{table}%

\subsubsection{Mechanistic models} \label{sec:mech}
Compartmental models are the standard biological, or mechanistic, approach for modeling infectious disease \cite{Keeling2007, Siettos2013, Lessler2016}.
Kermack and McKendrick proposed one such model, now known as the susceptible-infectious-recovered (SIR) model, in which members of a population transition through each compartment (susceptible to infectious to recovered) over the course of an epidemic \cite{Kermack1927}.
While this process mimics the behavior of an outbreak, the simplest model assumes that the population is ``well mixed'', such that each individual is equally likely to encounter any other individual.
Since this is unlikely, researchers can add more compartments (\eg adults and children), along with contact rates between compartments, or individually model each member of the population (\ie agent-based modeling) \cite{Eubank2004}.
Though greater complexity requires more modeling assumptions, compartmental models have been effective at estimating underlying infectious disease processes and the potential impact of interventions \cite{Lessler2016, Ferguson2006, Keeling2007, Reich2013}.

Compartmental models have been extended for use in forecasting infectious disease incidence.
Variations on the SIR model have been used to develop forecast models for influenza.\cite{Birrell2011, Nishiura2011, osthus2017forecasting}
Some of these approaches incorporate humidity into a SIRS compartmental model.\cite{Shaman2012, Shaman2013}
The Ross-MacDonald model, a compartmental model originally developed for malaria that accounts for interactions between mosquitoes and humans,\cite{Smith2012} has been used to make forecasts of  dengue fever.\cite{Dinh2016,Amaku2016,Zhu2018}
However, other mechanistic approaches to forecast dengue outbreaks have not modeled mosquitoes specifically.\cite{yamana2016superensemble}

\subsubsection{Classical statistical models}

On the statistical side of the modeling spectrum, many regression-style methods have been used for forecasting.
Perhaps the most well-known statistical method for time series is the auto-regressive integrated moving average, or ARIMA.\cite{Box1962}
ARIMA models use a linear, regression-type equation in which the predictors are lags of the dependent variable and/or lags of the forecast errors.
ARIMA and seasonal ARIMA (SARIMA) models are frequently applied to infectious disease time series \cite{Johansson2016, Siettos2013, Soyiri2013, Unkel2012, Ray2017}.
Lu \etalspace combined a SARIMA model with a Markov switching model (a type of compartmental model) to account for anomalies in the surveillance process \cite{Lu2010}.

Also under the subheading of trend and seasonal estimation are simple exponential smoothing strategies, known as Holt-Winters models \cite{holt2004forecasting, winters1960forecasting}.
Exponential smoothing techniques involve taking weighted averages of past observations with exponentially decreasing weights further from the present.
Holt-Winters in particular is known for its efficient and accurate predictive ability \cite{gelper2010robust, goodwin2010holt}.
These approaches have been used successfully in forecasting dengue fever \cite{buczak2018ensemble} and leprosy \cite{deiner2017short}.

Some researchers have used generalized linear regression models to develop infectious disease forecasts.
In some cases, researchers used lagged covariates (\eg temperature, rainfall, or prior incidence) to predict future incidence \cite{Haemig2011, Hii2012, Lowe2011, Moore2012,Reich2016a}.
Held and Paul also combined statistical and biological theory by building a regression model that consisted of three components of disease incidence: endemic, epidemic, and spatio-temporal epidemic (to account for spread of disease across locations) \cite{held2005statistical}.
This has become a well-established framework for forecasting infectious disease surveillance data \cite{Hohle2014,held2017probabilistic,Ray2017}, and is accompanied by open-source software implementing the methods \cite{meyer2017}. 

\subsubsection{Modern statistical methods}

Modern statistical methods, \ie not the classical time-series and regression-based approaches, are an increasingly popular way to forecast infectious disease incidence.
These methods include non-parametric approaches as well as more black-box machine-learning style algorithms.
We focus in this section on stand-alone forecasting methods, for a discussion on ensemble methods, see Section \ref{sec:ensembles}.

Statistical or machine learning approaches have been in existence for decades.
While machine-learning and statistical methods are sometimes classified separately\cite{Siettos2013}, we group them together as ``statistical'', as both terms encapsulate approaches that use patterns from past incidence in order to forecast future incidence.\cite{Myers2000}
These approaches can be used for `data mining', by which large amounts of data are extracted from various online sources for pattern-recognition tasks, or for modeling, using empirical methods such as random forests, neural networks, or or support vector machines that do not make any parametric model assumptions.
These techniques came about in the computer science and artificial intelligence communities (see, e.g. \cite{ho1995random}), but can also be expressed statistically \cite{Hastie2009}.

Several papers have found that machine-learning modeling methods can outperform standard statistical models for infectious disease forecasting: random forests outperformed ARIMA forecasting avian influenza \cite{Kane2014}, a maximum entropy model outperformed logistic regression forecasting hemorrhagic fever with renal syndrome \cite{Liu2014}, and fuzzy association rule mining outperformed logistic regression forecasting dengue outbreaks \cite{Buczak2012}.
Additionally, kernel conditional density estimation, a semi-parametric method, was shown to have more well-calibrated probabilistic forecasts than SARIMA and other regression-based approaches for forecasting highly variable dengue outbreaks in San Juan, Puerto Rico \cite{Ray2017}.
Heteroskedastic Gaussian processes showed better forecast accuracy than generalized linear models in forecasting dengue fever.\cite{johnson2018phenomenological}
Neural networks have also been used for forecasting influenza \cite{xu2017forecasting,wu2018deep}, Zika \cite{akhtar2018dynamic}, and Hepatitis A \cite{guan2004forecasting}.

\subsubsection{Comparisons between mechanistic and statistical models}

From an epidemiological perspective, mechanistic models have several clear advantages over statistical models. 
They are biologically motivated, and therefore have parameters that relate to well-established theory and can be interpreted by experts in the field.
They have been adapted in previous work to include vector dynamics, which could be important for forecasting outbreaks of vector-borne diseases.\cite{lourencco20142012,reiner2013systematic,jewell2015bayesian} 
Mechanistic models can flexibly incorporate features such as interventions or behavioral changes, which can be critical, especially if forecasts are desired for different intervention scenarios (see Section \ref{sec:challenges}).
While mechanistic models can be built to rely heavily on previously observed data, they also can be instantiated with very little prior data, such as in emerging outbreaks (see Section \ref{sec:emerging}).
Additionally, while forecasts from statistical models are typically bounded by trends that have been previously observed, mechanistic models can forecast outside of previously observed trends if the underlying states of the model call for such dynamics.

Despite these advantages, in forecasting settings where substantial historical data is available, statistical models may prove more effective at using past observed trends to forecast the future.
Many statistical models were designed to be either more flexibly or parsimoniously parameterized, meaning that they may be able to more easily capture dynamics common to infectious disease time-series such as auto-regressive correlation and seasonality.
Additionally, they can be built to rely less heavily on specific assumptions about a particular biological disease transmission model, giving them flexibility to adapt when the data does not follow a particular pattern. 
In other words, since any specified mechanistic model is necessarily a simplification of the true underlying disease process, the question is how much will forecast accuracy suffer as a result of the inevitable model misspecification. 
In many cases, heavily parameterized mechanistic models may be more sensitive to model misspecification than a more flexible statistical model.

In practice, however, the distinction between statistical and mechanistic models is not always sharply defined. 
Due to complexities of real-time surveillance systems, forecasting targets of public health interest often represent a mixture of different signals that would be challenging in practice to be forecasted accurately by a single mechanistic model. 
For example, surveillance data of case counts may include distorted signals due to figments of partially automated reporting systems. This might be manifested in decreased case counts during holidays, or an increase in case counts at the end of the season when final reports are being prepared.\cite{Reich2016a}
In other settings, the actual target of interest may be a composite measure, such as with influenza-like illness (ILI) in the US.\cite{reich2019collaborative}
In both of these settings, the signal which is being predicted may be driven by factors that are not directly relevant to trends in disease transmission (e.g. clinical visitation or reporting patterns, changes in the processes used for diagnosis or case reporting). 
In these settings, statistical models that can have a more flexible understanding of the trends they are using to fit data may be at an advantage over mechanistic models. 
Research has demonstrated the value of coupling flexible statistical formulations with mechanistic curve-fitting to produce accurate forecasts.\cite{Osthus2019,pei2017counteracting,funk2016real,asher2018forecasting} 

Despite many unanswered questions about when and in what settings one type of model will generally do better than the other, studies that make explicit, data-driven comparisons are fairly uncommon.
Multi-team infectious disease forecasting challenges provide some of the best data available on this important question.
For forecasting both seasonal influenza and dengue outbreaks, what limited data there are suggest that mechanistic and statistical approaches show fairly similar performance, with statistical models showing a slight advantage.\cite{reich2019collaborative,mcgowan2019collaborative,johansson2019advancing} 
A collaborative effort during the 2014 west Africa Ebola outbreak to forecast synthetic data showed fairly comparable results from mechanistic and statistical models and did not make an explicit comparison between the two.\cite{viboud2017rapidd}
Summary analyses from other challenges have not been published, but we note that a very simple quasi-mechanistic model was the best performing model in forecasting the pattern of emergence of Chikunguya in the Americas \cite{lega2016data, DARPA2015}.
In summary, more research is needed to improve our understanding about whether different types of forecasting methods can be shown to be more reliable than others, especially as data availability and resolution improves over time.

\subsection{Forecasting in emergent settings} \label{sec:emerging}

In emerging outbreak scenarios, where limited data is available, mechanistic models may be able to take advantage of assumptions about the underlying transmission process, enabling rudimentary forecasts even with minimal data.
On the other hand, many statistical models without assuming a mechanistic structure rely on past data to be able to make forecasts.
That said, any forecasts made in settings with limited data must be subjected to rigorous sensitivity analyses, as such forecasts will necessarily be heavily reliant on model assumptions.

A wide range of different mechanistic models have been used in settings where infectious disease forecasts are desired for an emerging threat. 
A simple non-linear growth model performed the best in a prospective challenge for forecasting Chikungunya in the Americas \cite{lega2016data}.
This growth model made the assumption that the rate of change in the cumulative number of cases through the course of an outbreak follows a parabolic rise and fall, by which it can quickly approximate the parameters of an SIR model and estimate the peak incidence, duration, and total number of cases during an epidemic.
Unlike many other mechanistic forecasting approaches, this model has a small number of parameters, is easy to fit, and makes only a few assumptions about the underlying disease process. 
A deterministic SEIR model was used to forecast synthetic Ebola epidemic data, showing comparable performance to other methods on the same data \cite{gaffey2017application, viboud2017rapidd}.
A stochastic SEIR model also forecasted synthetic Ebola data, and showed somewhat less reliable performance compared to other methods \cite{funk2016real}.
Data-driven agent-based models have also been shown to be a viable forecasting tool for emerging infectious diseases \cite{venkatramanan2017using}.
An SIR model, similar to one used to forecast seasonal dengue fever and influenza, was used with a more complex compartmental structure to forecast the spread of Ebola during the outbreak in West Africa in 2014 \cite{Shaman2014}.
A set of quasi-mechanistic models were used to forecast Zika virus transmission during 2017 to help plan for vaccine trials \cite{Asher2017}.

\subsection{Using external data sources to inform forecasts}

Traditional approaches to infectious disease forecasting often have relied on a single time-series, or multiple similar time-series (\eg incidence from multiple locations).
However, other types of epidemiological data may provide important information about current transmission patterns.

\subsubsection*{Vector data} \label{sub:vectordata}

The use of data on prevalence and abundance of disease vectors in forecasting models has not been extensively explored in the literature. 
This is likely due in part to climate data being more widely available and the belief that such data could serve as a good proxy for actual data on vectors.
That said, perhaps the most well-developed area for incorporating vector data into forecasts is the use of prevalence data in forecasting mosquito-borne diseases such as West Nile virus\cite{kilpatrick2013predicting,davis2017integrating,Defelice2017}, and dengue fever\cite{shi2015three}. 
However, none of these studies have explicitly quantified the added value of vector data on forecast accuracy.
While the hypothetical benefits of good vector surveillance data have been clearly quantified\cite{yamana2019framework}, the benefit of these data in practice (especially when other reasonable proxy data may be available; see ``Climate data'' below) is still unclear. 

\subsubsection*{Laboratory data} \label{sub:lab}

Leveraging laboratory data, collected either through passive or active surveillance strategies, may provide crucial data about what specific pathogens are currently being transmitted and could inform forecasting efforts.
This is an area that warrants more research, as few efforts have tackled the challenge of having laboratory test data inform forecasts at the population level. 
One model uses an aggregate measure of genetic distance of circulating influenza strains from the strains in the vaccine as a variable to help forecast peak timing and intensity of seasonal outbreaks in the US \cite{Du2017, Du2018}.
Some efforts have also been made to make strain-specific forecasts for influenza \cite{Kandula2017}.
Other efforts have focused on longer-term forecasts of what strains will predominate in a given season, with an eye towards providing information to influenza vaccine manufacturers \cite{Morris2017}.
These efforts have moved beyond influenza, and forecasting pathogen evolution is being worked on for a variety of different pathogens \cite{Hadfield2017}.

\subsubsection*{Expert opinion}
Another, and very different, kind of epidemiological data for forecasting is expert opinion. 
Long seen as a useful indicator in business applications \cite{Surowiecki2004}, expert opinion has recently begun to be used in infectious disease applications \cite{farrow2017human, deiner2017short}.
While not traditional clinical data, expert opinion surveys leverage powerful computers, \ie human brains, that can synthesize historical experience with real-time data.\cite{budescu2014identifying} 
Intuitive interfaces can facilitate the specification of quantitative and digitally entered forecasts from experts who need not be technically savvy, lowering the barriers to participation and subsequent analysis \cite{farrow2017human}.
In the 2016/2017 influenza season in the US, a forecast model based on expert opinion was a top-performer in a CDC-led forecasting competition.\cite{mcgowan2019collaborative,reich2019collaborative}
Human judgment and expert opinion surveys are a promising area for further forecasting research, especially in contexts with limited data availability.

\subsubsection*{Digital epidemiology}\label{sec:digiepi}
Digital epidemiology has been defined as the use of digital data for epidemiology when the data were ``not generated with the primary purpose of doing epidemiology'' \cite{Salathe2018}.
Broadly speaking, this might refer to online search query data, social media data, satellite imagery, or climate data, to name a few. 
These resources may hold promise for forecasters who want to incorporate ``Big Data'' streams into their models.
In the past 10 years, much research has explored the potential for leveraging multiple data streams to improve forecasting efforts, but this practice is still in its nascent stages. 
So far, the utility of digital epidemiological data for forecasting has been somewhat limited, perhaps due to challenges in our understanding of how digital data generated by human behavior and interactions with the digital world relate to epidemiological targets \cite{moran2016epidemic, Priedhorsky2017, Salathe2018}.

Perhaps the most famous and controversial example of using digital data streams to support infectious disease prediction surround the early promising performance\cite{Ginsberg2009, Dugas2012} and later dismal failure \cite{Lazer2014} of Google Flu trends to predict the influenza-like-illness in the US.  
Google Flu trends was based on tracking influenza-related search terms entered into the search engine.  
Although Google eventually discontinued the public face of the project due to poor performance, criticism of the Google Flu trends approach centered around how data was included or excluded, interpreted, and handled rather than the algorithm that produced the actual forecasts \cite{Santillana2014, Olson2013}.
Ongoing research on using search engine data in forecasting has continued despite the failure of Google Flu trends, producing incremental but consistent improvements to forecast accuracy \cite{Yang2017, Yang2015, McGough2017,Lu2018,osthus2019even}.
More focused search query data, such as data from clinician queries, has also been shown promise for assisting real-time forecasting efforts.\cite{santillana2014using,thorner2016correlation}

\subsubsection*{Climate data} \label{sub:climdata}

The use of climate data for epidemic forecasting serves as another clear example of re-purposing data for epidemiology. 
While climate factors are known biological drivers of infection risk (\eg the impact of absolute humidity on influenza virus fitness \cite{shaman2009absolute}, or temperature and humidity providing optimal conditions for mosquito breeding), the evidence supporting the use of climate data in forecasting models is mixed.
Climatological factors such as temperature, rainfall, and relative humidity were used to forecast annual counts of dengue hemorrhagic fever in provinces of Thailand \cite{lauer2018prospective}.
However, only temperature and rainfall were included after a rigorous covariate selection process and neither were included in the final model, although subanalyses showed variation in these associations across different geographic regions of Thailand.
Climate factors were shown to improve forecasts of dengue outbreak timing in Brazil\cite{Lowe2017}, but played a less influential role in dengue forecasts in Mexico \cite{Johansson2016}.
Aggregated measures of absolute humidity have been incorporated into influenza forecasts in the US \cite{Shaman2013, Yang2017}.
However, without clear standardization across these studies, these mixed results may reflect heterogeneity in the spatial and temporal scales at which forecasts are made, climate factors are measured and aggregated, and disease transmission actually occurs.

\subsection{Forecasting with ensembles} \label{sec:ensembles}

Ensemble forecasting models, or models that combine multiple forecasts into a single forecast, have been the industry standard in weather forecasting and a wide array of other prediction-focused fields for decades. 
By fusing together different modeling frameworks, ensembles that have a diverse library of models to choose from end up incorporating information from multiple different perspectives and sources.\cite{Hastie2009}
When appropriate methods are used to combine either point or probabilistic forecasts, the resulting ensemble should in theory always have better long-run performance than any single model.\cite{bates1969combination,makridakis1983averages,clemen2007aggregating}
However, researchers have suggested that adjustments are necessary to correct for the introduction of bias\cite{granger1984improved} and miscalibration\cite{gneiting2013combining} in the process of building ensemble models.

Ensembles have been increasingly used in infectious disease applications and have shown promising results.
For forecasting influenza, several model averaging approaches have shown improved performance over individual models \cite{Yamana2017,ray2018prediction,reich2019multimodel}.
Similar approaches have yielded similar results for dengue fever \cite{yamana2016superensemble}, lymphatic filariasis \cite{Smith2017}, and Ebola \cite{viboud2017rapidd}.

In many of these examples, however, the number and diversity of distinct modeling approaches was fairly small.
To unlock the full potential value of ensemble forecasting, as well as understanding the added value of contributions from new and different data sources or modeling strategies, more scalable frameworks for building forecast models are required. 
There is a need to develop infrastructure and frameworks that can facilitate the building of ensemble forecast models.\cite{george2019} 
This will require clear technical definitions of modeling and forecasting standards.
Given the history of improved forecast performance due to improvements in ensemble methodology in other fields that focus on non-temporal data, the prospects for continued ensemble development in the area of temporal and time-series data, and especially infectious disease settings, are bright.

\section{Components of a Forecasting System}

Due to the complex biological, social, and environmental mechanisms underlying infectious disease transmission, the true underlying processes that give rise to the observed data is unknown.
This makes choosing one model, or even multiple, from the selection of possible choices quite challenging.
How can one decide which model is the best for forecasting future targets?

Deciding on the structure of the forecasting exercise, including the targets and the evaluation metrics, provides critical information to help decide upon appropriate methods.
Because models perform differently depending on the forecast target, type of forecast, model training technique, and evaluation metric, it is important to specify the forecasting system prior to fitting the models.\cite{Armstrong2001} 
This ensures that the system is tailored to the particular design of the exercise.

\subsection{Forecast type}

When building a forecasting system, the first steps are to choose the forecast target (as described in Section \ref{sec:notation}) and type.
Forecast targets are often dictated by the goals of a public health initiative.
Researchers and public health officials collaborate to find a forecast target that is most useful for allocating resources and implementing interventions to reduce the severity of an infectious disease outbreak.

The forecast target helps inform the selection of the forecast type. Forecasts can typically be classified as either point forecasts or probabilistic forecasts. Some authors classify ``interval'' forecasts as a separate entity\cite{Diebold2001}, however, these can be seen as a simplified version of a fully probabilistic forecast.

A point forecast is a forecast of a single value that attempts to minimize a function of the error between that value and the eventually observed value.
The mean, median, or mode of a predictive distribution is often used as the point forecast for a specified target.
This choice may depend on the specific metric used for evaluation, as the mean is the theoretically optimal choice if point forecasts are being evaluated with mean squared error (MSE) and the median is optimal if being evaluated by mean absolute error (MAE).\cite{Reich2016} 
While point forecasts are simpler to produce and interpret, they may make simplifying assumptions about the underlying probability distribution, leading to low-quality forecasts.
For example, a point forecast based on the mean may represent a value for which there is actually a small likelihood of occurring if, for example, it lies between the peaks of a multi-modal distribution.
This could mislead officials and researchers into forecasting a medium-sized outbreak when the full distribution actually shows that the most likely future scenarios are for either low incidence or an epidemic outbreak.

Interval forecasts supplement point forecasts with a ``prediction interval'', or a range of likely values.
The nominal level of a prediction interval indicates the percentage of eventually-observed outcomes that should fall within that interval.
If a model makes 100 forecasts, about 95 should fall within the 95\% prediction interval.
More generally, a ($1-\alpha*100$)\% prediction interval can be thought of as the interval that has a significance level of $\alpha$.
Interval forecasts are typically derived from some form of a probabilistic model or assessment of in-sample forecast error or uncertainty. 

A fully probabilistic forecast must specify a probability distribution function
The goal of a probabilistic forecast is to assign the maximum probability to the true future value.
Probabilistic forecasts can specify a closed-form parametric density function (e.g., a Gaussian distribution with a mean and variance) or an empirical distribution, either with an empirical cumulative density function, a set of samples from the predictive density, or a binned density function, with probabilities assigned to a discrete set of possible outcomes.
Density estimation often requires simulation-generating methodology, which can be more time-consuming and computationally-intensive than other techniques.
Ongoing advances in computing continue to make density forecasting methods more feasible for researchers.
Density forecasts contain the most nuanced information of all of the forecasting methods, but are often the most difficult to interpret and communicate to non-expert collaborators.  

\subsection{Evaluation and scoring}
\label{subsec:eval}


There is a rich literature on scoring and evaluating all types of forecasts.
Of course appropriate metrics will depend on the forecasting setting and the scoring criteria for a particular exercise. 
In general, models should be fit with a loss function or ``goodness of fit'' criteria that that is similar or identical to the method that will be used to evaluate forecasts.

\subsubsection*{General principles for scoring forecasts}
Research suggests that metrics should be \textit{scale-independent}.\cite{Armstrong2001,Hyndman2006}
For example, within a single infectious disease time series, larger incidence values are both more difficult to forecast and often have larger errors on an absolute scale than smaller incidence values simply because they are larger numbers. 
Thus, incidence values near the seasonal peak are both larger and more variable than incidence near the seasonal nadir and, consequently, forecasting model error will depend on the size of the value it is forecasting.
In these situations, something closer to scale-independence can be achieved either by using logged metrics can weight errors more equally across different scales or by using relative measures of accuracy (see, e.g., equation \ref{eq:rMAE} below).\cite{Reich2016}

\textit{Metrics should be defined and finite in reasonable scenarios}.
This principle ensures that scores from single forecasts may be combined together, for example with an average.
If single forecasts could be infite in reasonable scenarios, one individual forecast could eclipse all other scores in a summary measure such as an average. 
However, even non-experts can agree that a model that forecasts negative values of disease incidence should be considered invalid and it could be appropriate to have an infinite scoring value for such a forecast.

Forecasts should be evaluated using \textit{proper scoring rules}.\cite{Gneiting2007}
A proper scoring rule incentivizes the forecaster to report exactly what the model predicts.
An improper scoring rule could incentivize adjusting the forecast reported by a model to ``game'' the system only to get a better score.
While using an improper scoring rule is unlikely to change the relative performance of a set of models dramatically, it can lead to settings where the best scoring forecast is one that has been modified in undesirable ways.
For example, when probabilistic forecasts for season peak week are scored by evaluating the probability assigned to the true peak week plus or minus one week (an improper score used by the US CDC in influenza forecasting competitions\cite{mcgowan2019collaborative,reich2019collaborative}), even high-scoring forecast models can be adjusted to have better scores by adjusting the probabilities assigned to different weeks in a systematic way that is `dishonest' to the original forecast.\cite{bracher2019}

Forecast accuracy should always be \textit{evaluated with out-of-sample observations and with as large of a sample of observations as is feasible}.
Since forecasts are by definition predictions of as-yet-unobserved data, it is critical to evaluate forecast accuracy on out-of-sample data.
This means that forecasts should only be evaluated on observations that were not used to fit or train the model.
Ideally, the out-of-sample data would be ``prospective'' in the sense that all of evaluated observations would be from a point in time after the training data.
However, in cases with limited data availability, this may not be feasible.
Data quantity can be a limiting factor for many real-world datasets, especially for infectious disease surveillance. 
The analyst must balance competing needs of having sufficient data for training a realistic model with holding out data (ideally prospectively) for cross-validation and testing.

\subsubsection{Evaluating point forecasts}

Point forecasts are typically evaluated on their own using metrics such as mean squared error (MSE) or the mean absolute error (MAE).
For comparative evaluation of point forecasts in practice, many researchers recommend using a metric that scales the forecasting error against that of a reference model \cite{Hyndman2006, Reich2016, Gneiting2007}.

One example is the relative mean absolute error (rMAE), which divides the mean absolute error of one forecasting model (model A) by the mean absolute error of a second model (model B): 
\begin{equation} 
\text{rMAE} =\frac{\sum_{t=1}^n|z_t-\hat{z}^{\text{A}}_t|}{\sum_{t=1}^n|z_t-\hat{z}^{\text{B}}_t|}. \label{eq:rMAE}
\end{equation}
Recall that $z_t$ is the target of interest forecasted at time $t$ and $\hat{z}^A_t$ represents the forecast from model A at time $t$.
In principle rMAE may be calculated between any two models, however, it is common for rMAE to be calculated for a set of models against a common `reference' or basline model in the denominator.

An additional desirable feature of rMAE is that it is interpretable for public health officials. When $rMAE<1$ this means that the forecasting model has less error than the reference model on the scale of the original data and $rMAE>1$ means that the forecasting model has more error than the reference model.
For example, if model A has an rMAE of 0.9 compared to a reference seasonal model of case counts for a particular disease that means that the predictions from model A were 10\% closer to the observed value than predictions from the reference model were.

\subsubsection{Evaluating interval forecasts}

Interval forecasts can be evaluated by their coverage rate and their width. 
Prediction intervals should be as narrow as possible while covering a proportion of forecasts approximately equal to that expected by its level.

Perhaps the most commonly used interval metric is the coverage rate (CR) which is simply the fraction of all $(1-\alpha)*100$\% prediction intervals that cover the true value. 
Therefore
$$ CR_\alpha = \frac{1}{T}\sum_{t=1}^T \mathbb{I}(l^\alpha_t \leq z_t \leq u^\alpha_t) $$
where $\mathbb{I}$ is the indicator function (equalling 0 if the expression inside is FALSE and 1 if TRUE), and $l^\alpha_t$ and $u^\alpha_t$ are the lower and upper bounds of a $(1-\alpha)*100$\% prediction interval for observation $z_t$.
The observed $CR_\alpha$ can be evaluated for its proximity to $(1-\alpha)$.

A interval evaluation metric recommended for it's being a `proper' scoring metric is\cite{Gneiting2007}

$$
S^{\text{int}}_\alpha (z_t, u^\alpha_t, l^\alpha_t) = \frac{1}{T}\sum_{t=1}^T (u^\alpha_t-l^\alpha_t) + \frac{2}{\alpha} (l^\alpha_t-z_t) \mathbb{I} (z_t<l^\alpha_t) + \frac{2}{\alpha} (z_t-u^\alpha_t) \mathbb{I} (z_t>u^\alpha_t)
$$
where it is desirable to minimize the score $S^{\text{int}}_\alpha$.
Forecasting models are penalized for having wider intervals and for having observed values that fall far outside of the intervals.
Observations that fall outside of large prediction intervals (small $\alpha$) are penalized more than those that fall outside of small prediction intervals (large $\alpha$).

\subsubsection{Evaluating probabilistic forecasts}

In the long run, probabilistic forecasts should have distributions that are consistent with the distribution of the observed values. 
Models that assign more weight to the eventually observed values should be scored better than those that do not.\cite{Gneiting2007}
A commonly used proper scoring rule for probabilistic forecasts is the log score. Aggregated across many predictions, this metric is defined as 
$$\text{LogS}=\frac{1}{T}\sum_{t=1}^T\log \hat p(z_t)$$
where $\hat p(\cdot)$ is the estimated probability of observing the target $z_t$.
However, this metric is sensitive to outliers, as any observation with a forecasted probability of zero causes the metric to go to negative infinity (though adjustments can be made to avoid this).
As an alternative, Funk \etalspace recommend using multiple metrics to evaluate the unbiasedness, calibration, and sharpness of infectious disease forecasts \cite{Funk2017}.

The continuous ranked probability score (CRPS) is a proper scoring rule that measures the difference between the forecasted and observed cumulative distributions \cite{Hersbach2000}.
This metric measures both the bias and the uncertainty of the forecasted density and thus rewards forecasts that assign weight closer to the observed value, even if it doesn't assign much weight exactly on the observed value.
A point forecast with no uncertainty will have a CRPS equal to the absolute error of the forecast.
Unbiased forecasts with more uncertainty will have a higher CRPS than for unbiased forecasts with less uncertainty, however biased forecasts with more uncertainty can have a smaller CRPS than biased forecasts with less uncertainty.
While CRPS is scale-dependent, dividing the CRPS of a forecasting model by the MAE of a benchmark model (as in the relative mean absolute error) yields a scale-independent continuous ranked probability skill score \cite{Bradley2011,Bogner2018}.

\subsubsection{Tests for comparing models}

Comparing the relative performance of different models can yield important insights about the benefits of a unique data source or one approach over another.
However, in a time-series forecasting context (the most common for infectious disease forecasts) several clear statistical challenges are present when making forecast comparisons.
Most importantly, outcomes and forecasts at multiple nearby timepoints will be correlated with each other, reducing and complicating the understanding of the power of these tests to detect ``significant'' differences. 
The Diebold-Mariano test is the most well-known test to compare the errors from two forecasting models.\cite{diebold2002comparing} 
This method is implemented in software packages, for example, the {\tt forecast} package in R.\cite{hyndman2019package,hyndman2008book}
Other permutation based approaches have also been used to compare the significance between the forecast errors for two models.\cite{ray2018prediction}

However, in the infectious disease forecasting literature it has not yet become common practice to run such tests. 
Instead, authors have tended to rely on simple numeric comparisons of forecast errors between two or more models.
Not running a formal test allows for the possibility that the observed differences are due to chance.
However, from a practical perspective, as long as the forecasts are truly prospective in nature and the comparisons presented were the only ones made, such a comparison can provide tangible information about which model to choose for decision-making.
In situations where a definitive statement about the predominance of one model over another is desired, a formal test will likely be the best evidence available.

\subsection{Model training and testing}

In order for a forecasting model to be useful for researchers or officials it needs to be generalizable to data beyond the observations that were used for fitting.
For instance, a model that perfectly forecasts monthly dengue incidence over the past ten years, but performs worse than a reasonable guess---\eg the average monthly incidence---over the next five years is not very useful.
We would be better off using the reasonable guess instead of the forecasting model.
Though we can never be certain that our best model will perform well outside of our dataset, we can get a better idea of its \textit{out-of-sample} performance using cross-validation with a training and testing set.
We illustrate this concept with an example from Lauer \etal \cite{lauer2018prospective}, in which we forecasted annual dengue hemorrhagic fever (DHF) incidence in Thailand for 76 provinces.

A central challenge for forecasting in general is to train a model in such a way that we minimize the error on the unobserved data, i.e., the test data.
For real-time forecasts, the test data will be unobserved at the time a model is specified.
When forecasting is performed retrospectively for data that has already been collected, the test data will already have been observed. 
Strictly speaking, such an experiment is not forecasting at all, as it does not involve making predictions about data that have not yet been observed.
Nonetheless, it can be an important part of forecasting research to understand model performance in these settings.
To ensure the validity and generalizability of such findings, it is critical to only use the test data once the models have been specified from the training phase.

Typically, forecasters use cross-validation methods to evaluate and estimate error on not-yet-seen observations.\cite{Hastie2009}
There is a rich literature on cross-validation methods, including some techniques specific to time-series applications.\cite{Bergmeir2018}
These methods tend to reward slightly more complex models that may have more error on the testing data than a smaller or simpler model would.\cite{Shao1993}
Thus, in addition to selecting the model that performs best in the training phase by a pre-specified information criterion or cross-validation metric, forecasters should also choose a more parsimonious model that has more error in the training phase as a check against overfitting.\cite{Ng1997}

\begin{figure}[htbp]
\begin{center}
\includegraphics[width=\textwidth]{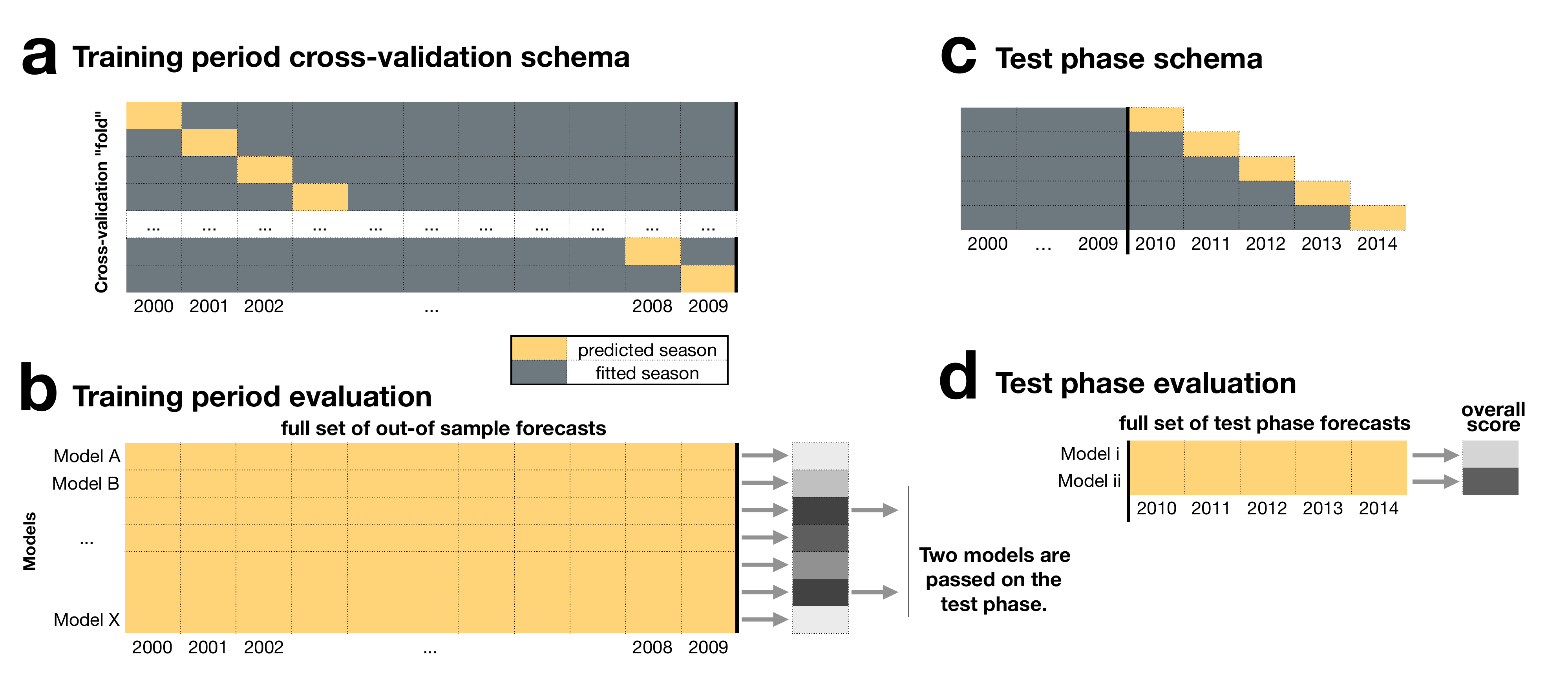}
\caption{A schematic showing the process of model training, cross-validation, and testing for making predictions of annual incidence of dengue hemorrhagic fever using province-level data from Thailand. 
(a) A diagram of the cross-validation experiment for predicting annual dengue incidence. One year is left out at a time, models are fit to data from the other seasons, and out-of-sample predictions are obtained for the left-out season. 
(b) A suite of models are fit in the training period, with a full set of out-of-sample forecasts obtained for each year. These scores are then summarized into a single summary out-of-sample, cross-validated score. Two models are chosen and passed into the testing phase.
(c) A diagram of the way in which forecasts are made for each of the five testing years. The data from all previous years are fit and the forecasts for the current year are made in a prospective fashion.
(d) Only the two models selected in the training phase are implemented in the test phase. The prospective forecasts across all five test-phase seasons are aggregated and summarized into a final overall score.
}
\label{fig:cv-schema}
\end{center}
\end{figure}

Prior to fitting any model, we split our data into a `training' sample (for initial model selection) and a `testing' sample (for final model evaluation) \cite{Stone1974, Hastie2009}.
These steps are standard practice in the field, and similar to formal recommendations for modeling disease surveillance data.\cite{althouse2015enhancing}
In this example, data from years 2000 through 2009 (760 observations) served as the training phase data and years 2010 through 2014 (380 observations) served as the test phase data (Figure \ref{fig:cv-schema}a).
The training sample is used for model experimentation and parameter tuning.  

There is no one right answer for how to split data into training and testing sets, however the choice may be informed by prior knowledge about the modeling setting.
We chose to model the training phase data using leave-one-year-out cross-validation, so that each year's training forecast would be conditional on the remaining 9 years of data. 
While this does not preserve strict ordering of data (e.g., the data from 2000 is predicted based on a model fit to data from 2001 through 2009), it ensures that each of the training period forecasts is based on the same amount of data.
The alternative would have been to implement a training regimen that would have predicted 2001 based only on 2000 data, 2002 based only on 2000-2001 data. 
Due to the limited length of this dataset, this would mean that early forecasts would be based on substantially less data.
Using leave-one-year-out cross-validation ensures that each of the 10 years of training forecasts will have the same amount of data, and a roughly similar amount of data that we expect to have in the test phase.
However, if substantially more data were available prior to 2000 (say, more than 5 years of data) then it might have been desirable to implement prospective cross-validation in the training phase as well.

\begin{figure}[htbp]
\begin{center}
\includegraphics[width=\textwidth]{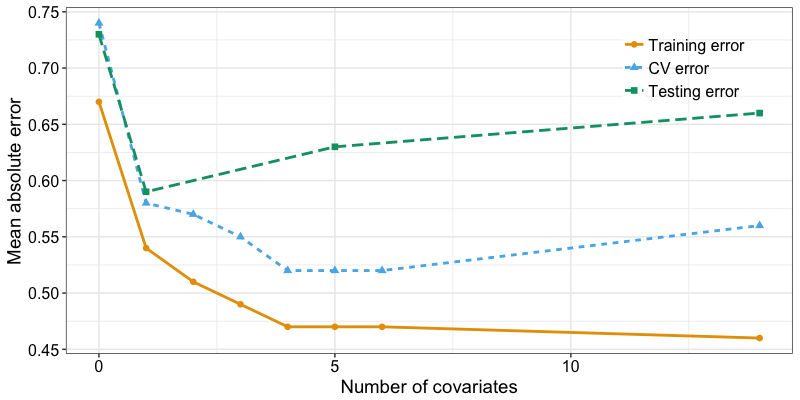}
\caption{
Error in forecasting annual incidence of dengue hemorrhagic fever in Thailand, across 76 provinces.
Mean absolute errors are measured as the absolute difference between the log observed incidence and the log predicted incidence, where lower values indicate better forecast accuracy.
The training phase in-sample error (orange, solid line), out-of-sample cross-validation (CV) error (blue, dotted), and testing phase error (green, dashed) plotted as a function of the number of model covariates ($x$-axis).
The training error monotonically decreases as the number of covariates increases.
The CV error is minimized at 5 covariates and better approximates the testing error than the training error, especially for fewer covariates.
The univariate model (1 covariate) had the least error in the testing phase.
}
\label{fig:mae}
\end{center}
\end{figure}

The training period is complete once all candidate models have generated out-of-sample forecasts for each of the training years.
Typically, a small number of models are selected to pass into the test phase.
In our example, we ran leave-one-year-out cross validation on the training phase data to select our model.
In this procedure, we fit a model on 9 of the 10 years to predict the final year --- \eg fitting on 2001-2009 to predict 2000.
We repeated this to predict the province-level DHF incidence in each of the 10 years, recorded the error for each prediction, and then took the mean absolute error across all predictions and called it the ``cross-validation (CV) error'' for a given model  (Figure \ref{fig:cv-schema}b).
We performed cross-validation for 202 models with different specifications and covariate combinations.
The model that minimized the CV error had 5 covariates, while the model that minimized the in-sample residual error across the entire training phase had 14 covariates (\autoref{fig:mae}).
In addition to the 5-covariate model, we also selected the smallest model within one standard deviation of the smallest CV error --- in this case it was a univariate model --- to forecast the test phase.

In the test phase, we implemented a prospective testing schema to more realistically simulate real-time forecasts (Figure \ref{fig:cv-schema}c).
This \textit{rolling-origin-recalibration} window, as it has been called\cite{Bergmeir2012}, is implemented by first fitting the model to the training data to forecast the first test phase observation. Then the first observation from the test phase is moved into the training data, the model is re-fit and the second test phase observation is forecasted.
We used this method in the testing phase of our example as it is good for evaluating how a model might perform in real-time, as more data is collected and assimilated into the model fitting process.
In our example, we evaluated the testing phase forecasts using relative mean absolute error (rMAE) of our model over baseline forecasts based on the ten-year median incidence rate for each province (Figure \ref{fig:cv-schema}d).
The univariate model had less testing phase error than the best cross-validation model (\autoref{fig:mae}). 
Additionally, the univariate model had about 20\% less error than a baseline forecast, a rolling 10-year-median for each province (data not shown).

When a forecaster is interpreting the results, she should recommend for future use the model that performed best in the test phase.
The goal at the outset of this exercise was to find the model that makes the best forecasts that are generalizable outside of our data, which is defined by the performance on the testing phase.
Specific times or places where a different model showed good results could be areas for future forecasting activities, however analysts should cautious about over-interpreting small-sample size results in training or test phase results.

Prior to splitting the data into training and testing, it is critical to think about how many observations are needed for each sample. 
There need to be enough training observations to properly fit the model and there need to be enough testing observations to properly evaluate the model.
With a short time-series, there may be too few data to split and thus only cross validation can be conducted on all of the data; in this scenario, interpretations about the model performance will be weaker than those with a separate testing phase.

\section{Operationalizing forecasts for public health}

Making forecasts on infectious disease data is difficult due to the culmination of a variety of factors, from the microscopic to the population level, which are difficult to do in any circumstances, but are compounded by logistical issues when trying to make forecasts in real time.
Logistical challenges include assimilating newly-collected data into the forecasting framework, accounting for delays in case reporting, and effectively communicating the model results to public health officials.

\subsection{Reporting delays} \label{sec:delays}

Making forecasts in real time introduces the dimension of reporting delays into our forecasting models.
From a disease surveillance perspective, reporting delays are a timeliness issue which varies by features of the disease (ease of diagnosis, incubation time), surveillance entity (local, state, or national government), transmission type (electronic or not), and case load, as well as variability in reporting between surveillance systems \cite{Jajosky2004, BonacicMarinovic2015}.
From a data perspective, this means that observed values in the recent past are subject to change.

Since most forecasting models make the assumption that values used for fitting are fixed, we need to adapt our forecasting process by ``nowcasting'' observed values in the recent past.
One method of now-casting is to only include ``sufficiently complete'' data up such that a forecasting model can make stable forecasts.
For instance, 75\% of dengue hemorrhagic fever cases in Thailand were reported to the Thai Ministry of Public Health within 10 weeks of infection \cite{Reich2016a}.
To account for this, we ignored the last 12 weeks (actually 6 biweeks, to be exact) before forecasting forward.
In our notation, we fit our model to data $y_{1:(t+k-1)}$ to make a k-step forecast, $y_{t+k}$, where $k=-6$.
Another method of nowcasting is to use past reporting delays to model recent incomplete counts.
Several frameworks have been proposed to nowcast infectious disease incidence based on past reporting rates.\cite{Hohle2014,bastos2019modelling}
Other approaches for nowcasting have incorporated digital surveillance data.\cite{brooks2018nonmechanistic,osthus2019even}

When case counts for prior time periods are subject to change, it is important for researchers to have a collection of data ``snapshots'', so that past situations can be investigated retrospectively with the information that was available at the time.
Thus, database of should contain records of cases as they are reported, containing the date of illness and incidence that is timestamped upon deposit into the database.

\subsection{Communication of results}

Public health authorities have shown increasing interest in working with infectious disease forecasters in the light of recent important public health crises.
Starting in 2009 with the pandemic influenza A outbreak, public health officials turned to forecasters for estimates of burden and burden averted due to vaccines and antivirals.  
During the Ebola outbreak in 2014, public health officials again turned to prediction for specific information regarding the potential outbreak size and intervention impacts.  
These efforts highlight how infectious disease forecasting can support public health practice now and in the future.

\subsubsection*{What makes a good forecast?}

Previous work in meteorology has outlined 3 distinct forecast attributes of a forecast that contribute to its usefulness, or ``goodness'' \cite{Murphy1993}.
If we apply these guidelines to infectious disease forecasting, we can surmise that a forecast is good if it is (a) \textit{consistent}: reflecting the forecaster's best judgment, (b) \textit{quality}: forecasts conditions that are actually observed during the time being forecasted, and (c) \textit{valuable}: informs policy or other decision-making that results in increased benefits to individuals or society.  

For a forecast to reflect the forecaster's ``best judgment'' means that the forecast is reasonable based on the forecaster's expert knowledge base, prior experience, and best and current methodology.
The forecaster's internal judgments are not usually available for evaluation or quantification, but could say that a forecast is not a reflection of best judgment if we discover that a forecasting model contains an error or under some conditions produces values outside the range of possible values.

To meet the conditions for high quality, forecasted values must correspond closely to observed values.
The field of forecast verification is so vast and specialized that we could not possibly give it a comprehensive treatment here.
Suffice it to say that reducing error is central goal of the field of forecasting.
Examples of quality measurement approaches include the mean absolute error and the mean-squared error, which reflect forecast accuracy.
Other examples include measures of bias, skill (often a comparison to reference models), and uncertainty \cite{Jolliffe2003}.

Infectious disease forecasts are valuable if they are used to influence decisions.
Sometimes value can sometimes be accessed in quantitative units (\eg lives or money saved or lost).
Forecast quality influences value to a large extent, but so do other more qualitative features of how the forecast is communicated.
For example, a forecast will have a larger impact on decision-making if it is timely, presented clearly, and uses meaningful units in addition to being accurate or improving on a previous system.




\section{Conclusion and Future Directions}

There has been a great deal of progress made in infectious disease forecasting, however the field is very much still in its infancy. 
Forecasts of epidemics can inform public health response and decision-making, including risk communication to the general public, and timing and spatial targeting of interventions (\eg vaccination campaigns or vector control measures).
However, to maximize the impact that forecasts can have on the practice of public health, interdisciplinary teams must come together to tackle a variety of challenges, from the technological and statistical, to the biological and behavioral.
To this end, the field of infectious disease forecasting should emphasize the development and integration of new theoretical frameworks that can be directly linked to tangible public health strategies.

To facilitate the development of scalable forecasting infrastructure and continued research on improving forecasting, the field should focus on developing data standards for both surveillance data and forecasts themselves. 
This will foster continued methodological development and facilitate scientific inquiry by enabling standard comparisons across forecasting efforts. 
One key barrier to entry to this field is that the problems are operationally complex: a model may be asked to forecast multiple targets at multiple different times, using only available data at a given time.
Converging on standard language and terminology to describe these challenges is key to growing the field and will accelerate discovery and innovation for years to come.

\bibliographystyle{unsrt}
 \newcommand{\noop}[1]{}

\end{document}